\documentclass[iop]{emulateapj}

\def\ltsima{$\; \buildrel < \over \sim \;$}
\def\simlt{\lower.5ex\hbox{\ltsima}}
\def\gtsima{$\; \buildrel > \over \sim \;$}
\def\simgt{\lower.5ex\hbox{\gtsima}}
\def\kpc{{\rm\,kpc}}

\def\pc{{\rm\,pc}}

\def\deg{^\circ}

\def\s{\ifmmode \widetilde \else \~\fi}
\def\={\overline}

\def\spose#1{\hbox to 0pt{#1\hss}}

\def\lta{\mathrel{\spose{\lower 3pt\hbox{$\mathchar"218$}}
     \raise 2.0pt\hbox{$\mathchar"13C$}}}
\def\gta{\mathrel{\spose{\lower 3pt\hbox{$\mathchar"218$}}
     \raise 2.0pt\hbox{$\mathchar"13E$}}}
\def\Dt{\spose{\raise 1.5ex\hbox{\hskip3pt$\mathchar"201$}}}    
\def\dt{\spose{\raise 1.0ex\hbox{\hskip2pt$\mathchar"201$}}}    

\def\dotsfill{\leaders\hbox to 1em{\hss.\hss}\hfill}

\def\FeH{{\rm[Fe/H]}}

\shorttitle{Lacerta~I and Cassiopeia~III}
\shortauthors{N. F. Martin et al.}

\begin{document}


\title{Lacerta~I and Cassiopeia~III: Two luminous and distant Andromeda satellite dwarf galaxies found in the $3\pi$ Pan-STARRS1 survey}


\author{Nicolas F. Martin$^{1,2}$, Colin T. Slater$^{3}$, Edward F. Schlafly$^2$, Eric Morganson$^2$, Hans-Walter Rix$^2$, Eric F. Bell$^3$, Benjamin P. M. Laevens$^{1,2}$, Edouard J. Bernard$^4$, Annette M. N. Ferguson$^4$, Douglas P. Finkbeiner$^5$, William S. Burgett$^6$, Kenneth C. Chambers$^6$, Klaus W. Hodapp$^6$, Nicholas Kaiser$^6$, Rolf-Peter Kudritzki$^6$, Eugene A. Magnier$^6$, Jeffrey S. Morgan$^6$, Paul A. Price$^7$, John L. Tonry$^6$, Richard J. Wainscoat$^6$}
\email{nicolas.martin@astro.unistra.fr}

\altaffiltext{1}{Observatoire astronomique de Strasbourg, Universit\'e de Strasbourg, CNRS, UMR 7550, 11 rue de l'Universit\'e, F-67000 Strasbourg, France}
\altaffiltext{2}{Max-Planck-Institut f\"ur Astronomie, K\"onigstuhl 17, D-69117 Heidelberg, Germany}
\altaffiltext{3}{Department of Astronomy, University of Michigan, 500 Church St., Ann Arbor, MI 48109}
\altaffiltext{4}{Institute for Astronomy, University of Edinburgh, Royal Observatory, Blackford Hill, Edinburgh EH9 3HJ, UK}
\altaffiltext{5}{Harvard-Smithsonian Center for Astrophysics, 60 Garden Street, Cambridge, MA 02138, USA}
\altaffiltext{6}{Institute for Astronomy, University of Hawaii at Manoa, Honolulu, HI 96822, USA}
\altaffiltext{7}{Department of Astrophysical Sciences, Princeton University, Princeton, NJ 08544, USA}

\begin{abstract}
We report the discovery of two new  dwarf galaxies, Lacerta~I/Andromeda~XXXI (Lac~I/And~XXXI) and Cassiopeia~III/Andromeda~XXXII (Cas~III/And~XXXII), in stacked Pan-STARRS1 $r_\mathrm{P1}$- and $i_\mathrm{P1}$-band imaging data. Both are luminous systems ($M_V \sim -12$) located at projected distances of $20.3\deg$ and $10.5\deg$ from M31. Lac~I and Cas~III are likely satellites of the Andromeda galaxy with heliocentric distances of $756^{+44}_{-28}\kpc$ and $772^{+61}_{-56}\kpc$, respectively, and corresponding M31-centric distances of $275\pm7\kpc$ and $144^{+6}_{-4}\kpc$. The brightest of recent Local Group member discoveries, these two new dwarf galaxies owe their late discovery to their large sizes ($r_h = 4.2^{+0.4}_{-0.5}$\,arcmin or $912^{+124}_{-93}\pc$ for Lac~I; $r_h = 6.5^{+1.2}_{-1.0}$\,arcmin or $1456\pm267\pc$ for Cas~III), and consequently low surface brightness ($\mu_0\sim26.0$~mag/arcsec$^2$), as well as to the lack of a systematic survey of regions at large radii from M31, close to the Galactic plane. This latter limitation is now alleviated by the $3\pi$ Pan-STARRS1 survey, which could lead to the discovery of other distant Andromeda satellite dwarf galaxies.
\end{abstract}


\section{Introduction}

The salience of satellite dwarf galaxies for understanding galaxy formation in a cosmological context has been made all the more evident in the past decade with the discovery of numerous faint Local Group galaxies. These faint systems are not only important to understand the faint end of galaxy formation (e.g. \citealt{koposov09,revaz12,vargas13}) but also their distribution around their host can test the hierarchical formation induced by the favored cosmological paradigm (e.g. \citealt{pawlowski12,ibata13a}). The Andromeda satellite system is arguably the ideal system to study in order to tackle these questions as its proximity renders it easily observable, while our outsider's perspective facilitates its complete mapping, provided one has access to high-quality, multi-color, and wide-field photometry.

The number of known Andromeda satellite dwarf galaxies has soared over the last decade, thanks mainly to two large photometric surveys of this region of the sky. The Sloan Digital Sky Survey (SDSS) enabled the discovery of a handful of relatively bright ($M_V\simlt-8.5$) systems, initially from a dedicated stripe along the major axis of the Andromeda galaxy (And~IX, And~X; \citealt{zucker04,zucker07}), which was later complemented by a more systematic coverage of parts of the region around M31 in Data Release~8 (And~XXVIII, And~XXIX; \citealt{slater11,bell11}). The Pan-Andromeda Archaeological Survey (PAndAS) is, on the other hand, a survey dedicated to the study of the stellar populations within the halo of the Andromeda galaxy, out to distances of $\sim150\kpc$. Alone, it has enabled the discovery of 16 unambiguous new Andromeda satellite dwarf galaxies \citep{martin06b,martin09,ibata07,mcconnachie08,richardson11} with total magnitudes ranging from $M_V\simeq-6.0$ to $M_V\simeq-10.0$.

The great success of these surveys in unveiling the M31 satellite system also highlights their coverage limitations. And~VI, VII \& XXVIII are all examples of dwarf galaxies at projected distances of at least about $30\deg$ (corresponding to $\sim 400\kpc$ M31-centric distances), yet SDSS has only surveyed regions that stay well away from the Milky Way disk, mostly south of M31, and PAndAS is limited to a cone of $\sim10\deg$ from M31. The Panoramic Survey Telescope and Rapid Response System~1 (Pan-STARRS1; \citealt{kaiser10}) is a photometry survey that does not have these limitations and, with final stacked data reaching slightly deeper than the SDSS \citep{metcalfe13}, it has the potential to provide a much more complete understanding of the M31 dwarf galaxy satellite system out to and beyond the virial radius of Andromeda ($\sim300\kpc$).

In this paper, we present the discovery of the two dwarf galaxies found in stacked Pan-STARRS1 imaging data. As they are located in the Lacerta and Cassiopeia constellations, but are also likely Andromeda satellites, we follow the dual-naming convention emphasized in \citet{martin09} to avoid ambiguity. We therefore dub the first dwarf galaxy both Lacerta~I and Andromeda~XXXI (Lac~I/And~XXXI) and the second one both Cassiopeia~III and Andromeda~XXXII (Cas~III/And~XXXII). The data set used for the discovery is summarized in \S2, while the derivation of the dwarf galaxies' properties is described in \S3, before we conclude in \S4.

\section{The Pan-STARRS1 survey}

The PS1 $3\pi$ survey (Chambers et al., in preparation) is a systematic imaging survey of the sky north of $\delta=-30\deg$ in five optical and near-infrared photometric bands ($g_\mathrm{P1}r_\mathrm{P1}i_\mathrm{P1}z_\mathrm{P1}y_\mathrm{P1}$; \citealt{tonry12}). Conducted on a dedicated 1.8m telescope located on Haleakala, in Hawaii, the system relies on a  1.4-Gpixel, $3.3\deg$-field-of-view camera for rapid mapping of the sky. Any location in the survey is observed repeatedly for a planned four times per year per filter, conditions permitting, with exposure times of $43/40/45/30/30$ seconds in the $g_\mathrm{P1}/r_\mathrm{P1}/i_\mathrm{P1}/z_\mathrm{P1}/y_\mathrm{P1}$-bands, respectively \citep{metcalfe13}. The median seeing values in these bands are $1.27/1.16/1.11/1.06/1.01$ arcseconds. Images are automatically processed through the Image Processing Pipeline \citep{magnier06,magnier07,magnier08} to produce the stacked data catalogue.

At the time of this paper, the region around Andromeda had been observed for three consecutive seasons from May 2010, with up to twelve exposures per filter. Chip gaps, poor observing conditions, and technical problems, however, mean that a varying number of exposures are available for stacking at any given location. For point-sources, this leads to $10\sigma$-completeness limits\footnote{These detection limits correspond to the median magnitude of local stars with photometric uncertainties close to 0.1 magnitudes.} of 22.1 in the  $r_\mathrm{P1}$ band, and 22.2 in the  $i_\mathrm{P1}$ band for the surroundings of Lac~I. Around Cas~III, these limits are 22.0 and 21.9, respectively. Checking the spatial completeness from the presence of 2MASS stars, bright enough to be observed with PS1, yields an area coverage factor $> 95$ percent in either band.

Flags were used to remove spurious detections that are above the saturation limit, blended with other sources, thought to be defect, saturated pixels, or are considered significant cosmic rays or extended sources. Sources for which either the point-spread function fit is bad, the moments, the local sky, or the local sky variance could not be measured were also removed from the data set. Finally, detections for which the peak lands on a diffraction spike, a ghost, a glint, or near a chip edge were discarded as well, in addition to sources with signal-to-noise below 5.

The photometric calibration of the stacked PS1 data is still under development, with spurious offsets in the photometry for subsets of detections. In the region of Lac~I, a comparison of the stacked photometry with that of the single exposure PS1 observations, calibrated at the 0.01-magnitude level by \citet{schlafly12}, yields no significant offset in the $i_\mathrm{P1}$ band, whereas the $r_\mathrm{P1}$ band has $\sim20$ percent of sources offset by $\sim0.2$\,magnitudes. In the Cas~III surroundings, a small fraction of measurements are affected at the 0.05 level in the $r_\mathrm{P1}$ band, and at the 0.1 level in the $i_\mathrm{P1}$ band. However, since the majority of point sources have a reliable photometry, we have decided to nevertheless keep theses discrepant sources in the catalogue.

Throughout, the distance to M31 is assumed to be $779^{+19}_{-18}\kpc$ and the uncertainties on this values are modeled by drawing directly from the posterior probability distribution function provided by \citet{aconn12}. All magnitudes are dereddened using the \citet{schlegel98} maps, with the extinction coefficients for the PS1 filters provided by \citet{schlafly11}: $A_{g_{P1}}/E(B-V) = 3.172$, $A_{r_{P1}}/E(B-V) = 2.271$, and $A_{i_{P1}}/E(B-V) = 1.682$.

\section{Lacerta~I and Cassiopeia~III}
\begin{figure}
\begin{center}
\includegraphics[width=0.92\hsize,angle=270]{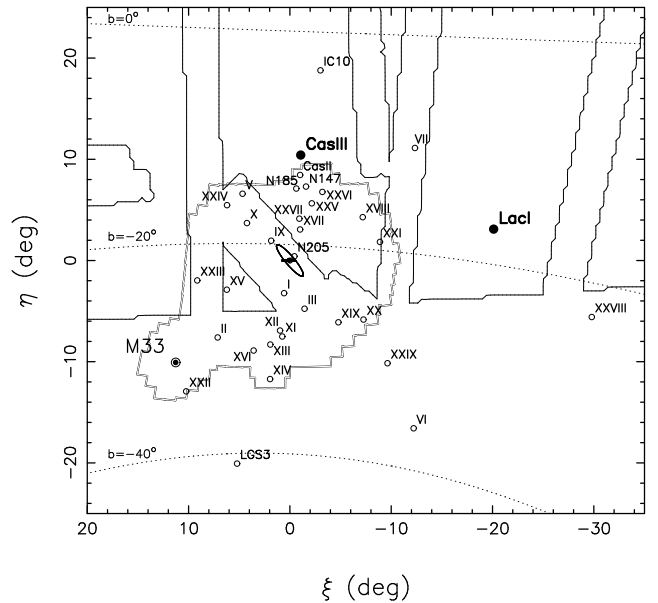}
\caption{\label{map} Map of dwarf galaxies located in the surroundings of M31 and M33, projected on the tangential plane centered on M31. Known satellites are shown as hollow circles while the two new discoveries, Lac~I and Cas~III, are represented by filled circles. The limits of the SDSS and PAndAS are represented by the thin-line and the double-line polygons, respectively. The dotted lines highlight lines of constant Galactic latitude at $b=0\deg$, $-20\deg$, and $-40\deg$ from top to bottom.}
\end{center}
\end{figure}

In a first pass at searching for unknown M31 satellites within $30\deg$ from their host, a map was produced from the catalogue sources selected in the color-magnitude diagram (CMD) to loosely correspond to potential red giant branch (RGB) stars at the distance of Andromeda ($0.0\simlt (r_\mathrm{P1}-i_\mathrm{P1})_0\simlt0.8$, $i_{\mathrm{P1},0}<20.8$). This map showed a handful of high-significance peaks, including the well-known M31 dwarf galaxies And~I, II, III, V, VI, and LGS3. The CMD of significant peaks which did not overlap with known satellites were inspected by eye, leading to the clear discovery of two compact stellar system located at ICRS $(\alpha,\delta)=(22^\mathrm{h}58^\mathrm{m}16.3^\mathrm{s},+41\deg17'28'')$ and $(\alpha,\delta)=(0^\mathrm{h}35^\mathrm{m}59.4^\mathrm{s},+51\deg33'35'')$ that have no counterparts in NED or Simbad. We designate them as Lac~I/And~XXXI and Cas~III/And~XXXII and show the location of the two dwarf galaxies comparatively to the M31 satellite dwarf galaxy system in Figure~\ref{map}.

\begin{figure*}
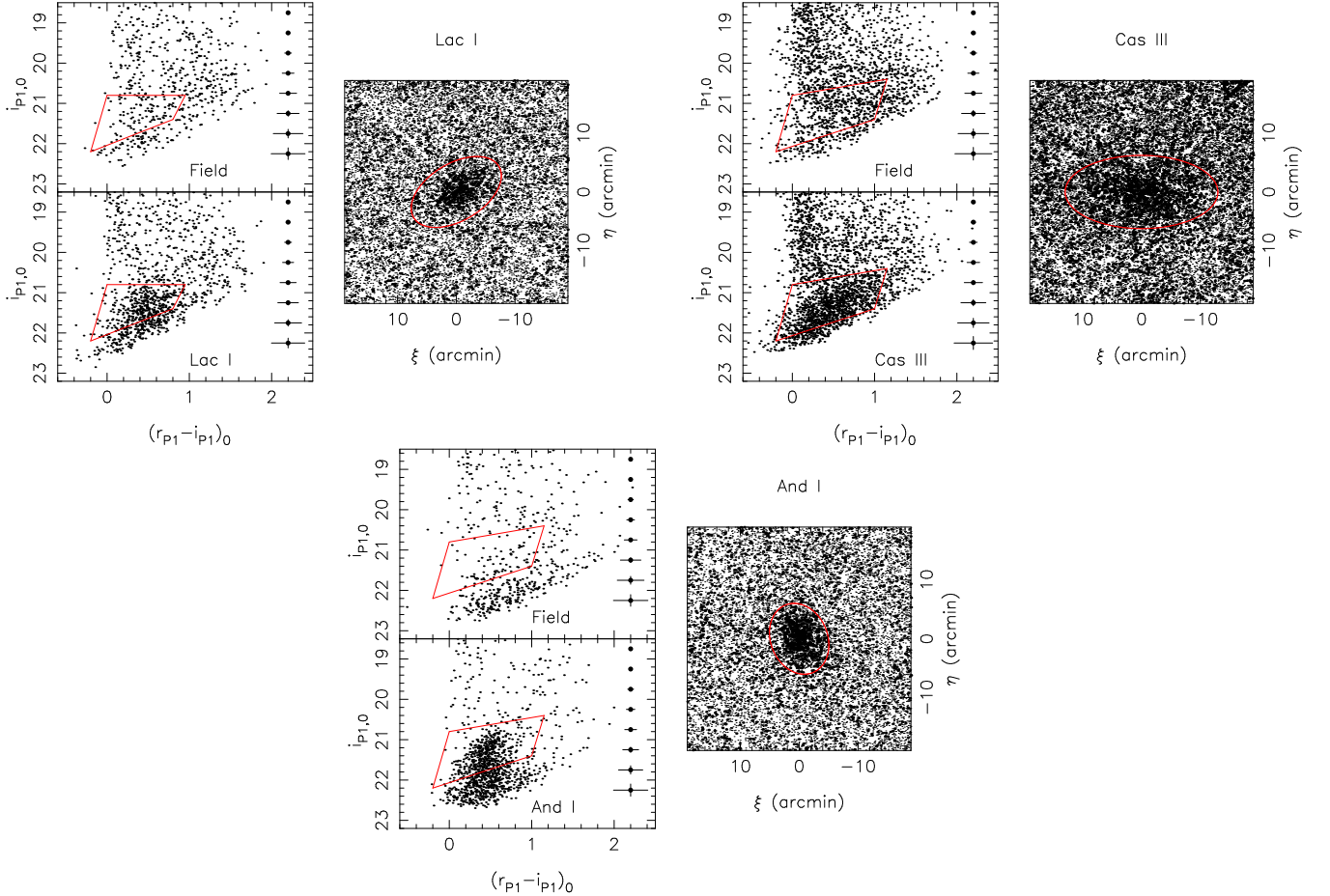

\begin{center}
\includegraphics[width=0.34\hsize,angle=270]{Fig2a.ps}
\hspace{0.8cm}
\includegraphics[width=0.34\hsize,angle=270]{Fig2b.ps}
\includegraphics[width=0.34\hsize,angle=270]{Fig2c.ps}
\caption{\label{CMD_map}\emph{Top, left-hand panels:} CMD of the region within two half-light radii of Lac~I (bottom-left) and that of a field region of the same coverage (top-left), along with the map of PS1 sources around Lac~I (right). The dwarf galaxy's CMD exhibits an overdensity of stars in the region expected for bright RGB stars at the distance of M31, matched by a spatial overdensity in the map. For the CMDs, the error-bars on the right-hand side of the panels represent the average photometric uncertainties and the red polygon delineate the selection box that was used to isolate candidate Lac~I stars, shown as large dots in the spatial map (other sources are shown as small dots). The red ellipse overlaid on the map delineates the region within two half-light radii of the dwarf galaxy, according to the favored structural parameters determined in \S~\ref{structure}. \emph{Top, right-hand panels:} Similar plots for Cas~III. \emph{Bottom panels:} Similar plots for And~I. The And~I structural parameters are taken from \citet{mcconnachie12}.}
\end{center}
\end{figure*}

The top, left-hand panels of Figure~\ref{CMD_map} present the CMD of stars within two half-light radii of Lac~I (see below for the estimation of its structural parameters); for comparison, the panels also include the CMD of a field region with the same coverage, but offset by a degree towards the south-east. Lac~I produces a clear overdensity of stars within the RGB color-selection, typical of stars close to the tip of the red giant branch (TRGB) at the distance of M31. Isolating these stars yields the map shown to the right of the CMDs and reveals that they are also clumped in a clear spatial overdensity, with a size of a few arcminutes, typical of M31 satellite dwarf galaxies. The same is true for Cas~III (top, right-hand panels), despite a more severe contamination from the prominent Milky Way disk.

For comparison, the bottom panels of Figure~\ref{CMD_map} show the same set of plots for the well-known Andromeda~I dwarf galaxy, also present in the stacked PS1 data, albeit in a region of deeper data. The CMDs of both And~I and its background region are less populated owing to the lower Galactic foreground contamination ($b=-16.7\deg$ for Lac~I and $b=-11.2\deg$ for Cas~III vs. $b=-24.8\deg$ for And~I), but otherwise displays a similar overdensity of stars when compared to the two discoveries.

\begin{table}
\caption{\label{properties}Derived properties of Lac~I/And~XXXI and Cas~III/And~XXXII}
\begin{tabular}{l|cc}
 & Lac~I & Cas~III\\
\hline
$\alpha$ (ICRS) & $22^\mathrm{h}58^\mathrm{m}16.3^\mathrm{s}$& $0^\mathrm{h}35^\mathrm{m}59.4^\mathrm{s}$\\
$\delta$ (ICRS) & $+41\deg17'28''$ & $+51\deg33'35''$\\
$\ell$ ($\deg$) & 101.1 & 120.5\\
$b$ ($\deg$) & -16.7 & -11.2\\
$E(B-V)$\footnote{From \citet{schlegel98}.} & 0.133 & 0.193\\
$(m-M)_0$ & $24.40\pm0.12$ & $24.45\pm0.14$\\
Heliocentric distance (kpc) & $756^{+44}_{-28}$ & $772^{+61}_{-56}$\\
M31-centric distance (kpc) & $275\pm7$ & $144^{+6}_{-4}$\\
$M_V$ & $-11.7\pm0.7$ & $-12.3\pm0.7$\\
$\mu_0$ (mag/arcsec$^2$) & $25.8\pm0.8$ & $26.4\pm0.8$ \\
Ellipticity & $0.43\pm0.07$ & $0.50\pm0.09$\\
Position angle (N to E; $\deg$) & $-59\pm6$ & $-90\pm7$\\
$r_h$ (arcmin) & $4.2^{+0.4}_{-0.5}$ & $6.5^{+1.2}_{-1.0}$\\
$r_h$ (pc) & $912^{+124}_{-93}$ & $1456\pm267$\\
\end{tabular}
\end{table}

\subsection{Structure}
\label{structure}
\begin{figure*}
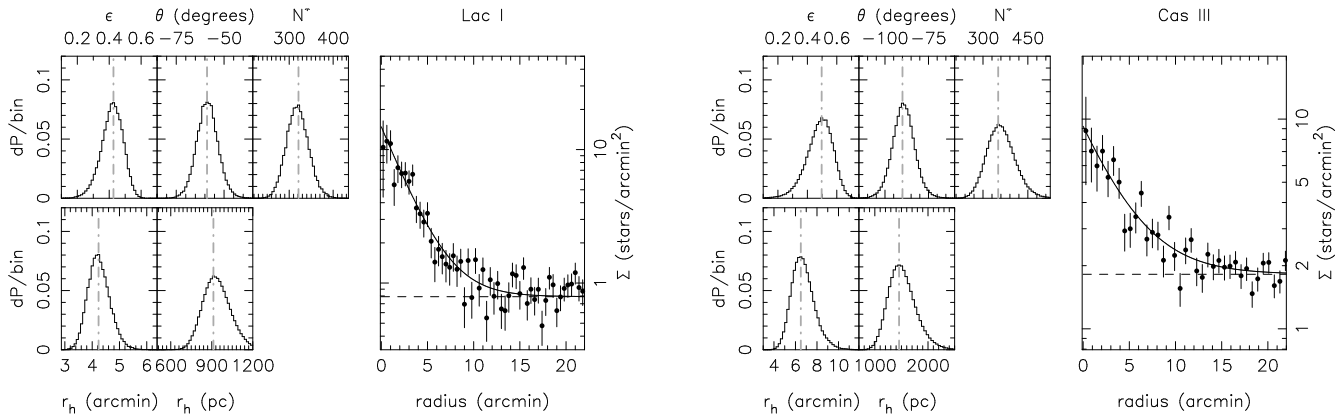

\begin{center}
\includegraphics[width=0.30\hsize,angle=270]{Fig3a.ps}
\hspace{0.8cm}
\includegraphics[width=0.30\hsize,angle=270]{Fig3b.ps}
\caption{\label{AndXXXI_struct}\emph{Left-hand panels:} The likelihood distribution functions for the structural parameters of Lac~I, marginalized over other parameters. The gray dash-dotted lines correspond to the maximum of the distributions, whose parameter values are taken as the favored model parameters. The right-most panel of this set displays the binned radial density profile of the dwarf galaxy built by assuming the favored centroid, ellipticity, and position angle (filled circles), compared to the favored exponential profile and background (full line). The dashed line corresponds to the favored estimation of the background. The uncertainties on counts are computed assuming Poisson uncertainties. \emph{Right-hand panels:} Similar plots for Cas~III.}
\end{center}
\end{figure*}

The structural parameters of Lac~I and Cas~III are determined following the maximum likelihood technique presented by \citet{martin08b}, and updated in Martin et al. (2013, in prep) to allow for a full Markov Chain Monte Carlo estimate of the parameters, allowing for missing data. For systems that contain relatively few stars, a likelihood-based estimate of the structural parameters is far less biased and much more robust than estimates that rely on binning the data \citep{martin08b}. In this instance, the algorithm explores the likelihood of a family of exponential models with a constant background level to reproduce the distribution of PS1 stars around Lac~I or Cas~III. The parameters of the dwarf galaxy models are its centroid, its exponential half-light radius ($r_h$), the position angle of its major axis ($\theta$), its ellipticity ($\epsilon = 1-b/a$, with $a$ and $b$ being, respectively, the major and minor axis scale lengths of the system), and the number of member stars within the chosen CMD selection box ($N^*$), which is tied to the background level through the number of stars in the data set. Uncertainties on these parameters are determined from the one dimensional, marginalized likelihood distribution functions and the location at which the likelihood reaches $e^{-0.5}\simeq0.61$ of the peak value. For a Gaussian distribution, the corresponding parameter values bound the $1\sigma$ confidence interval.

The likelihood distribution functions for these parameters, constrained on the PS1 data of the region within $25'$ of the centroid of Lac~I or Cas~III, are presented in Figure~\ref{AndXXXI_struct} for the most relevant parameters. These are well constrained and, in particular, the high significance of $N^*$ in both cases leaves no ambiguity to the existence of the new dwarf galaxies. Their radial density profiles are also displayed in the same figure and, for both dwarf galaxy, compares the data, binned following the most likely centroid, ellipticity, and position angle, with the exponential model of most likely half-light radius and number of stars. In both cases, the very good agreement between the two is testimony to the quality of the fit.

The measured structure of Lac~I and Cas~III are rather typical of bright Andromeda satellite galaxies with moderate ellipticities ($0.42\pm0.07$ and $0.50\pm0.09$, repectively). The size of the two systems is on the high side of the size distribution of M31 satellite dwarf galaxies: $r_h = 4.2^{+0.4}_{-0.5}$\,arcmin or $912^{+124}_{-93}\pc$ for Lac~I and $r_h = 6.5^{+1.2}_{-1.0}$\,arcmin or $1456\pm267\pc$ for Cas~III (see below for their heliocentric distance estimates). 

The properties of Lac~I and Cas~III are summarized in Table~\ref{properties} and the MCMC chains are available on demand to the corresponding author.

\subsection{Distance}
\begin{figure*}
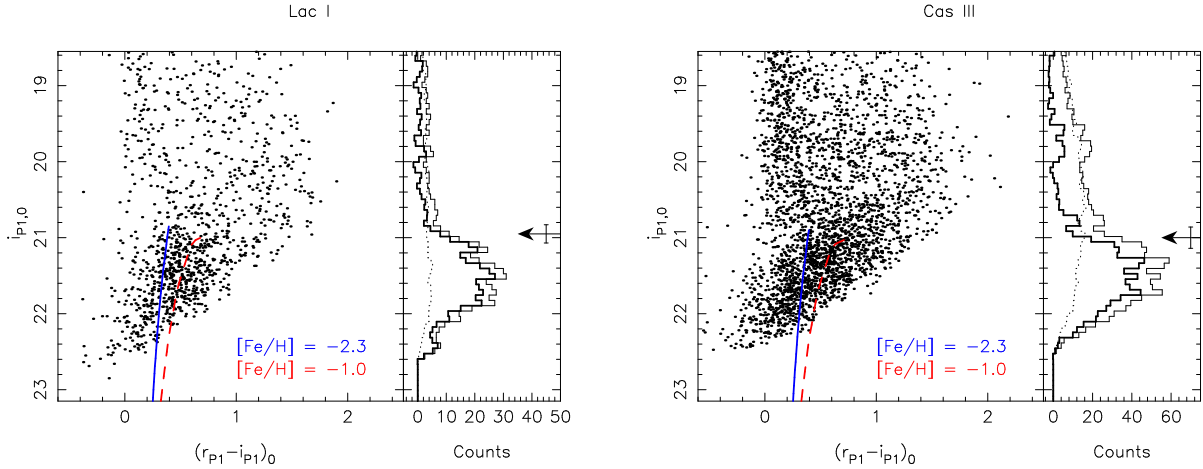

\begin{center}
\includegraphics[width=0.34\hsize,angle=270]{Fig4a.ps}
\hspace{0.8cm}
\includegraphics[width=0.34\hsize,angle=270]{Fig4b.ps}
\caption{\label{AndXXXI_LF}\emph{Left-hand panels:} CMD of stars within two half-light radii of Lac~I, with \textsc{Parsec} isochrones of age 12.7 Gyr and metallicity $\FeH=-2.3$ and $-1.0$ overlaid as a full blue line and a dashed red line, respectively. The binned luminosity function of the CMD stars is represented by a thin line in the right panel, while that of a local field region scaled to the same coverage is represented by a dotted line, and the resulting background-corrected luminosity function of Lac~I is displayed with a thick line. The estimated location of the TRGB is indicated by the arrow, and its uncertainty is represented by the vertical error bar attached to it. \emph{Right-hand panels:} Similar plots for Cas~III.}
\end{center}
\end{figure*}

The magnitude of the TRGB is a very good distance indicator in the case of Andromeda dwarf galaxies (e.g., \citealt{aconn11}), especially in the absence of deep photometric data. The tip's location also has the benefit of being relatively insensitive to the metallicity and age of stellar populations, as long as these are reasonably old and metal-poor, a condition that is fulfilled for all known M31 dwarf spheroidal galaxies. We apply the likelihood technique implemented by \citet{slater11} to determine the distance to And~XXVIII, later confirmed to be accurate from deeper observations of this system (Slater et al., in preparation). The TRGB of Lac~I is estimated at $i_{\mathrm{P1},0}^\mathrm{TRGB} = 20.95\pm0.07$, and that of Cas~III at $i_{\mathrm{P1},0}^\mathrm{TRGB} = 21.00\pm0.10$. In both cases, the measurement is in good agreement with the dwarf galaxy's luminosity function, as shown in Figure~\ref{AndXXXI_LF}.

The TRGB magnitude has been estimated at $M_{i,\mathrm{SDSS}}^\mathrm{TRGB}=-3.44\pm0.10$ by \citet{bellazzini08b}. Its conversion onto the PS1 photometric system using the color equations of \citet{tonry12} yields $M_{i,\mathrm{P1}}^\mathrm{TRGB}=-3.45\pm0.10$ and allows us to calculate the distance modulus of Lac~I: $(m-M)_0 = 24.40\pm0.12$ . This distance modulus corresponds to a heliocentric distance of $756^{+44}_{-28}\kpc$ and, when also accounting for the M31 distance uncertainty, an M31-centric distance of $275\pm7\kpc$\footnote{The low uncertainty on the M31-centric distance stems from Lac~I being close to the point of closest distance to M31 along the line of sight, i.e. the geometric tangent point, irrespective of the moderate uncertainty on its heliocentric distance.}. Given the dearth of isolated Local Group dwarf galaxies, it still appears that likely Lac~I is an Andromeda satellite despite its large M31-centric distance; its properties are similar to other remote satellites like And~VI, VII, XXVIII, or XXIX, which have now all been confirmed as Andromeda satellites from radial velocity measurements \citep{mcconnachie12,tollerud13}.

In the case of Cas~III, the distance modulus is $(m-M)_0 = 24.45\pm0.14$, leading to a heliocentric distance of $772^{+61}_{-56}\kpc$, and an M31-centric distance of $144^{+6}_{-4}\kpc$. At this distance, Cas~III is even more likely to be a member of the Andromeda satellite system. Of course, radial velocities are needed to add evidence that both Lac~I and Cas~III are truly bound to M31.

\subsection{Total magnitude}
To estimate the total magnitude of the two dwarf galaxies, we first sum up the flux contribution of all stars within their half-light radius that are in the interval $20.7<i_{\mathrm{P1},0}<21.5$ and also broadly fall in the RGB's color range. The contaminating flux from foreground sources is calculated in the same way from a large local field region, later scaled to the area within the half-light radius of Lac~I and Cas~III, and subtracted from the flux of the dwarf galaxies. We correct for unobserved stars by integrating a metal-poor ($\FeH=-1.7$) and old (12.7\,Gyr) \textsc{Parsec} luminosity function in the PS1 photometric system \citep{bressan12}. This yields the corrected apparent magnitude: $m_{i\mathrm{P1},0} = 12.8$ and $m_{r\mathrm{P1},0} = 13.3$ for Lac~I and $m_{i\mathrm{P1},0} = 12.1$ and $m_{r\mathrm{P1},0} = 12.6$ for Cas~III. Data are not deep enough in $g_{\mathrm{P1}}$ to yield a reliable estimate of the apparent magnitude in that band, complicating the transformation to the widely used $V$-band magnitude.

In order to circumvent this difficulty, we rely on a differential measurement with the $r_\mathrm{P1} and i_\mathrm{P1}$ magnitudes of And~I, II, III and V as measured from PS1 data, in the PS1 bandpasses. We repeat the above exercise for these four dwarf galaxies, account for their distance modulus difference with those of Lac~I and Cas~III, and tie the total magnitude of the two discoveries to that of these well-known dwarf galaxies, yielding $M_V = -11.7\pm0.7$ for Lac~I and $M_V = -12.3\pm0.7$ for Cas~III. The crudeness of this method and the multitude of assumptions it relies on are made evident by the large uncertainties, measured as the dispersion of the magnitude derivations from the four reference dwarf galaxies. Nevertheless, the CMDs of the two new dwarf galaxies apparently containing a similar number of stars to that of And~I  (Figure~\ref{CMD_map}) acknowledges the fact that they have a total magnitude similar to that of And~I ($M_V=-11.7\pm0.1$).

\section{Final remarks}
We report the discovery of two new dwarf galaxies, Lac~I/And~XXXI and Cas~III/And~XXXII, from stacked PS1 imaging data. At a heliocentric distance of $756^{+44}_{-28}\kpc$, Lac~I is located in the environs of the Andromeda galaxy, at an M31-centric distance of $275\pm7\kpc$, making it a likely satellite of Andromeda. Cas~III is located closer to M31 in projection and, combined with our measurement of its heliocentric distance ($772^{+61}_{-56}\kpc$), is well within the M31 virial radius with an M31-centric distance of only $144^{+6}_{-4}\kpc$. The membership of both dwarf galaxies to the M31 satellite system shall be confirmed kinematically through our ongoing spectroscopic program of both Lac~I and Cas~III. The discovery of Cas~III is of particular interest as it is located on the rotating disk of dwarf galaxies that includes about half of Andromeda's dwarf galaxies \citep{aconn13,ibata13a}.

Both dwarf galaxies are quite large ($r_h=912^{+124}_{-93}\pc$ and $r_h = 1456\pm267\pc$), yet this is not unexpected for dwarf galaxies or their brightness. \citet{brasseur11b} estimated that, for a total magnitude of $M_V = -12.0$, Andromeda dwarf galaxies have an average size of $775\pc$, with a lognormal dispersion of 0.23, implying that Lac~I is barely above the average size and Cas~III just slightly more than one dispersion away. One also has to keep in mind that, of course, it is the largest systems that are the most likely to have so far avoided detection, biasing new discoveries towards larger systems than were discovered from photographic surveys.

\begin{figure}
\begin{center}
\includegraphics[width=\hsize]{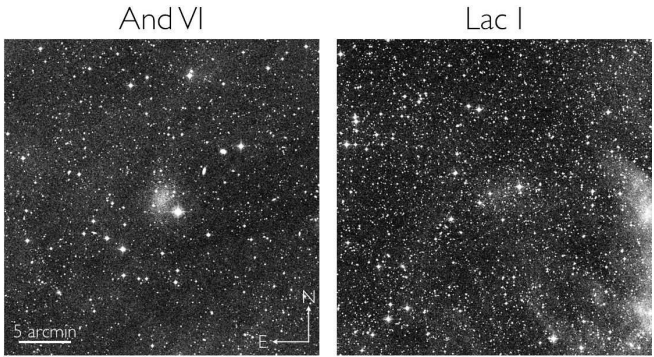}
\caption{\label{POSS_Images} POSS~II F (Red) image of And~VI (left) and Lac~I (right). The images are 30 arcmin on the side, centered on the two dwarf galaxies, with north to the top and east to the left. And~VI is readily visible while Lac~I appears as a faint amount of unresolved light. More significant wispy features due to the foreground Milky Way dust appear at the edge of the Lac~I image.}
\end{center}
\end{figure}

To a certain degree, the only unexpected property of Lac~I and Cas~III are their total magnitude, $M_V=-11.7\pm0.7$ and $-12.3\pm0.7$, respectively, which are an order of magnitude more luminous than any member of the cohort of Local Group dwarf galaxies discovered in the last decade. Although surprising at first, one must remember that, prior to PS1, the region of the two discoveries had only been systematically observed in the optical by photographic surveys. Regardless, one might wonder why such bright dwarf galaxies were not discovered earlier and, in particular, why the searches based on photographic plates that led to the discovery of, e.g., the fainter And~VI \citep{armandroff99,karachentsev99}, did not also pick up Lac~I. The explanation lies in the large half-light radii of the new systems, which are 2--3 times larger than that of And~VI, leading to their much fainter surface brightness ($\mu_0 \sim 26$~mag/arcsec$^2$). In fact, if one is to look meticulously at the location of Lac~I in the POSS plates, one can see a very faint fuzz of unresolved light that betrays the presence of the dwarf galaxy (Figure~\ref{POSS_Images}). The signal of this detection was likely too easily confused with the wispy Milky Way dust filaments, evermore present at this latitude, and disregarded at the time. Cas~III, with its even larger size, does not appear as a significant detection in the POSS plates.

Providing a detailed assessment of the stellar populations present in Lac~I and Cas~III is beyond the PS1 data and the scope of this discovery paper, even though we can nevertheless notice that the dwarf galaxies do not appear to harbor a population of very young and blue stars ($\simlt200$\,Myr). 

The strong signal of Lac~I and Cas~III in the PS1 data, along with the detectability of fainter systems such as And~XXIX, highlights the ability of the PS1 $3\pi$ stacked survey to expand our knowledge of the Andromedan satellite system and, beyond, that of the 1-Mpc heliocentric sphere. The new observations should prove particularly interesting in the region around M31 not observed by the deeper PAndAS, nor by the SDSS that shied away from the close-by Milky Way plane.

\acknowledgments

NFM gratefully acknowledges the CNRS for support through PICS project PICS06183. NFM \& HWR acknowledge support by the DFG through the SFB 881. CTS and EFB acknowledge support from NSF grant AST 1008342.

The Pan-STARRS1 Surveys (PS1) have been made possible through contributions of the Institute for Astronomy, the University of Hawaii, the Pan-STARRS Project Office, the Max-Planck Society and its participating institutes, the Max Planck Institute for Astronomy, Heidelberg and the Max Planck Institute for Extraterrestrial Physics, Garching, The Johns Hopkins University, Durham University, the University of Edinburgh, Queen's University Belfast, the Harvard-Smithsonian Center for Astrophysics, the Las Cumbres Observatory Global Telescope Network Incorporated, the National Central University of Taiwan, the Space Telescope Science Institute, the National Aeronautics and Space Administration under Grant No. NNX08AR22G issued through the Planetary Science Division of the NASA Science Mission Directorate, the National Science Foundation under Grant No. AST-1238877, and the University of Maryland.


\begin{thebibliography}{35}
\expandafter\ifx\csname natexlab\endcsname\relax\def\natexlab#1{#1}\fi

\bibitem[{{Armandroff} {et~al.}(1999){Armandroff}, {Jacoby}, \&
  {Davies}}]{armandroff99}
{Armandroff}, T.~E., {Jacoby}, G.~H., \& {Davies}, J.~E. 1999, \aj, 118, 1220

\bibitem[{{Bell} {et~al.}(2011){Bell}, {Slater}, \& {Martin}}]{bell11}
{Bell}, E.~F., {Slater}, C.~T., \& {Martin}, N.~F. 2011, \apjl, 742, L15

\bibitem[{{Bellazzini}(2008)}]{bellazzini08b}
{Bellazzini}, M. 2008, \memsai, 79, 440

\bibitem[{{Brasseur} {et~al.}(2011){Brasseur}, {Martin}, {Macci{\`o}}, {Rix},
  \& {Kang}}]{brasseur11b}
{Brasseur}, C.~M., {Martin}, N.~F., {Macci{\`o}}, A.~V., {Rix}, H.-W., \&
  {Kang}, X. 2011, \apj, 743, 179

\bibitem[{{Bressan} {et~al.}(2012){Bressan}, {Marigo}, {Girardi}, {Salasnich},
  {Dal Cero}, {Rubele}, \& {Nanni}}]{bressan12}
{Bressan}, A., {Marigo}, P., {Girardi}, L., {Salasnich}, B., {Dal Cero}, C.,
  {Rubele}, S., \& {Nanni}, A. 2012, \mnras, 427, 127

\bibitem[{{Conn} {et~al.}(2012){Conn}, {Ibata}, {Lewis}, {Parker}, {Zucker},
  {Martin}, {McConnachie}, {Irwin}, {Tanvir}, {Fardal}, {Ferguson}, {Chapman},
  \& {Valls-Gabaud}}]{aconn12}
{Conn}, A.~R., {Ibata}, R.~A., {Lewis}, G.~F., {Parker}, Q.~A., {Zucker},
  D.~B., {Martin}, N.~F., {McConnachie}, A.~W., {Irwin}, M.~J., {Tanvir}, N.,
  {Fardal}, M.~A., {Ferguson}, A.~M.~N., {Chapman}, S.~C., \& {Valls-Gabaud},
  D. 2012, \apj, 758, 11

\bibitem[{{Conn} {et~al.}(2011){Conn}, {Lewis}, {Ibata}, {Parker}, {Zucker},
  {McConnachie}, {Martin}, {Irwin}, {Tanvir}, {Fardal}, \&
  {Ferguson}}]{aconn11}
{Conn}, A.~R., {Lewis}, G.~F., {Ibata}, R.~A., {Parker}, Q.~A., {Zucker},
  D.~B., {McConnachie}, A.~W., {Martin}, N.~F., {Irwin}, M.~J., {Tanvir}, N.,
  {Fardal}, M.~A., \& {Ferguson}, A.~M.~N. 2011, \apj, 740, 69

\bibitem[{{Conn} {et~al.}(2013){Conn}, {Lewis}, {Ibata}, {Parker}, {Zucker},
  {McConnachie}, {Martin}, {Valls-Gabaud}, {Tanvir}, {Irwin}, {Ferguson}, \&
  {Chapman}}]{aconn13}
{Conn}, A.~R., {Lewis}, G.~F., {Ibata}, R.~A., {Parker}, Q.~A., {Zucker},
  D.~B., {McConnachie}, A.~W., {Martin}, N.~F., {Valls-Gabaud}, D., {Tanvir},
  N., {Irwin}, M.~J., {Ferguson}, A.~M.~N., \& {Chapman}, S.~C. 2013, \apj,
  766, 120

\bibitem[{{Ibata} {et~al.}(2007){Ibata}, {Martin}, {Irwin}, {Chapman},
  {Ferguson}, {Lewis}, \& {McConnachie}}]{ibata07}
{Ibata}, R., {Martin}, N.~F., {Irwin}, M., {Chapman}, S., {Ferguson}, A.~M.~N.,
  {Lewis}, G.~F., \& {McConnachie}, A.~W. 2007, \apj, 671, 1591

\bibitem[{{Ibata} {et~al.}(2013){Ibata}, {Lewis}, {Conn}, {Irwin},
  {McConnachie}, {Chapman}, {Collins}, {Fardal}, {Ferguson}, {Ibata}, {Mackey},
  {Martin}, {Navarro}, {Rich}, {Valls-Gabaud}, \& {Widrow}}]{ibata13a}
{Ibata}, R.~A., {Lewis}, G.~F., {Conn}, A.~R., {Irwin}, M.~J., {McConnachie},
  A.~W., {Chapman}, S.~C., {Collins}, M.~L., {Fardal}, M., {Ferguson},
  A.~M.~N., {Ibata}, N.~G., {Mackey}, A.~D., {Martin}, N.~F., {Navarro}, J.,
  {Rich}, R.~M., {Valls-Gabaud}, D., \& {Widrow}, L.~M. 2013, \nat, 493, 62

\bibitem[{{Kaiser} {et~al.}(2010){Kaiser}, {Burgett}, {Chambers}, {Denneau},
  {Heasley}, {Jedicke}, {Magnier}, {Morgan}, {Onaka}, \& {Tonry}}]{kaiser10}
{Kaiser}, N., {Burgett}, W., {Chambers}, K., {Denneau}, L., {Heasley}, J.,
  {Jedicke}, R., {Magnier}, E., {Morgan}, J., {Onaka}, P., \& {Tonry}, J. 2010,
  in Society of Photo-Optical Instrumentation Engineers (SPIE) Conference
  Series, Vol. 7733, Society of Photo-Optical Instrumentation Engineers (SPIE)
  Conference Series

\bibitem[{{Karachentsev} \& {Karachentseva}(1999)}]{karachentsev99}
{Karachentsev}, I.~D., \& {Karachentseva}, V.~E. 1999, \aap, 341, 355

\bibitem[{{Koposov} {et~al.}(2009){Koposov}, {Yoo}, {Rix}, {Weinberg},
  {Macci{\`o}}, \& {Escud{\'e}}}]{koposov09}
{Koposov}, S.~E., {Yoo}, J., {Rix}, H.-W., {Weinberg}, D.~H., {Macci{\`o}},
  A.~V., \& {Escud{\'e}}, J.~M. 2009, \apj, 696, 2179

\bibitem[{{Magnier}(2006)}]{magnier06}
{Magnier}, E. 2006, in The Advanced Maui Optical and Space Surveillance
  Technologies Conference

\bibitem[{{Magnier}(2007)}]{magnier07}
{Magnier}, E. 2007, in Astronomical Society of the Pacific Conference Series,
  Vol. 364, The Future of Photometric, Spectrophotometric and Polarimetric
  Standardization, ed. C.~{Sterken}, 153

\bibitem[{{Magnier} {et~al.}(2008){Magnier}, {Liu}, {Monet}, \&
  {Chambers}}]{magnier08}
{Magnier}, E.~A., {Liu}, M., {Monet}, D.~G., \& {Chambers}, K.~C. 2008, in IAU
  Symposium, Vol. 248, IAU Symposium, ed. W.~J. {Jin}, I.~{Platais}, \&
  M.~A.~C. {Perryman}, 553--559

\bibitem[{{Martin} {et~al.}(2008){Martin}, {de Jong}, \& {Rix}}]{martin08b}
{Martin}, N.~F., {de Jong}, J.~T.~A., \& {Rix}, H.-W. 2008, \apj, 684, 1075

\bibitem[{{Martin} {et~al.}(2006){Martin}, {Ibata}, {Irwin}, {Chapman},
  {Lewis}, {Ferguson}, {Tanvir}, \& {McConnachie}}]{martin06b}
{Martin}, N.~F., {Ibata}, R.~A., {Irwin}, M.~J., {Chapman}, S., {Lewis}, G.~F.,
  {Ferguson}, A.~M.~N., {Tanvir}, N., \& {McConnachie}, A.~W. 2006, \mnras,
  371, 1983

\bibitem[{{Martin} {et~al.}(2009){Martin}, {McConnachie}, {Irwin}, {Widrow},
  {Ferguson}, {Ibata}, {Dubinski}, {Babul}, {Chapman}, {Fardal}, {Lewis},
  {Navarro}, \& {Rich}}]{martin09}
{Martin}, N.~F., {McConnachie}, A.~W., {Irwin}, M., {Widrow}, L.~M.,
  {Ferguson}, A.~M.~N., {Ibata}, R.~A., {Dubinski}, J., {Babul}, A., {Chapman},
  S., {Fardal}, M., {Lewis}, G.~F., {Navarro}, J., \& {Rich}, R.~M. 2009, \apj,
  705, 758

\bibitem[{{McConnachie}(2012)}]{mcconnachie12}
{McConnachie}, A.~W. 2012, \aj, 144, 4

\bibitem[{{McConnachie} {et~al.}(2008){McConnachie}, {Huxor}, {Martin},
  {Irwin}, {Chapman}, {Fahlman}, {Ferguson}, {Ibata}, {Lewis}, {Richer}, \&
  {Tanvir}}]{mcconnachie08}
{McConnachie}, A.~W., {Huxor}, A., {Martin}, N.~F., {Irwin}, M.~J., {Chapman},
  S.~C., {Fahlman}, G., {Ferguson}, A.~M.~N., {Ibata}, R.~A., {Lewis}, G.~F.,
  {Richer}, H., \& {Tanvir}, N.~R. 2008, \apj, 688, 1009

\bibitem[{Metcalfe {et~al.}(2013)}]{metcalfe13}
Metcalfe, N., {et~al.} 2013, MNRAS, submitted

\bibitem[{{Pawlowski} {et~al.}(2012){Pawlowski}, {Kroupa}, {Angus}, {de Boer},
  {Famaey}, \& {Hensler}}]{pawlowski12}
{Pawlowski}, M.~S., {Kroupa}, P., {Angus}, G., {de Boer}, K.~S., {Famaey}, B.,
  \& {Hensler}, G. 2012, \mnras, 424, 80

\bibitem[{{Revaz} \& {Jablonka}(2012)}]{revaz12}
{Revaz}, Y., \& {Jablonka}, P. 2012, \aap, 538, A82

\bibitem[{{Richardson} {et~al.}(2011){Richardson}, {Irwin}, {McConnachie},
  {Martin}, {Dotter}, {Ferguson}, {Ibata}, {Chapman}, {Lewis}, {Tanvir}, \&
  {Rich}}]{richardson11}
{Richardson}, J.~C., {Irwin}, M.~J., {McConnachie}, A.~W., {Martin}, N.~F.,
  {Dotter}, A.~L., {Ferguson}, A.~M.~N., {Ibata}, R.~A., {Chapman}, S.~C.,
  {Lewis}, G.~F., {Tanvir}, N.~R., \& {Rich}, R.~M. 2011, \apj, 732, 76

\bibitem[{{Schlafly} \& {Finkbeiner}(2011)}]{schlafly11}
{Schlafly}, E.~F., \& {Finkbeiner}, D.~P. 2011, \apj, 737, 103

\bibitem[{{Schlafly} {et~al.}(2012){Schlafly}, {Finkbeiner}, {Juri{\'c}},
  {Magnier}, {Burgett}, {Chambers}, {Grav}, {Hodapp}, {Kaiser}, {Kudritzki},
  {Martin}, {Morgan}, {Price}, {Rix}, {Stubbs}, {Tonry}, \&
  {Wainscoat}}]{schlafly12}
{Schlafly}, E.~F., {Finkbeiner}, D.~P., {Juri{\'c}}, M., {Magnier}, E.~A.,
  {Burgett}, W.~S., {Chambers}, K.~C., {Grav}, T., {Hodapp}, K.~W., {Kaiser},
  N., {Kudritzki}, R.-P., {Martin}, N.~F., {Morgan}, J.~S., {Price}, P.~A.,
  {Rix}, H.-W., {Stubbs}, C.~W., {Tonry}, J.~L., \& {Wainscoat}, R.~J. 2012,
  \apj, 756, 158

\bibitem[{{Schlegel} {et~al.}(1998){Schlegel}, {Finkbeiner}, \&
  {Davis}}]{schlegel98}
{Schlegel}, D.~J., {Finkbeiner}, D.~P., \& {Davis}, M. 1998, \apj, 500, 525

\bibitem[{{Slater} {et~al.}(2011){Slater}, {Bell}, \& {Martin}}]{slater11}
{Slater}, C.~T., {Bell}, E.~F., \& {Martin}, N.~F. 2011, \apjl, 742, L14

\bibitem[{{Tollerud} {et~al.}(2013){Tollerud}, {Geha}, {Vargas}, \&
  {Bullock}}]{tollerud13}
{Tollerud}, E.~J., {Geha}, M.~C., {Vargas}, L.~C., \& {Bullock}, J.~S. 2013,
  ArXiv e-prints, ArXiv:1302.0848

\bibitem[{{Tonry} {et~al.}(2012){Tonry}, {Stubbs}, {Lykke}, {Doherty},
  {Shivvers}, {Burgett}, {Chambers}, {Hodapp}, {Kaiser}, {Kudritzki},
  {Magnier}, {Morgan}, {Price}, \& {Wainscoat}}]{tonry12}
{Tonry}, J.~L., {Stubbs}, C.~W., {Lykke}, K.~R., {Doherty}, P., {Shivvers},
  I.~S., {Burgett}, W.~S., {Chambers}, K.~C., {Hodapp}, K.~W., {Kaiser}, N.,
  {Kudritzki}, R.-P., {Magnier}, E.~A., {Morgan}, J.~S., {Price}, P.~A., \&
  {Wainscoat}, R.~J. 2012, \apj, 750, 99

\bibitem[{{Vargas} {et~al.}(2013){Vargas}, {Geha}, {Kirby}, \&
  {Simon}}]{vargas13}
{Vargas}, L.~C., {Geha}, M., {Kirby}, E.~N., \& {Simon}, J.~D. 2013, ArXiv
  e-prints, ArXiv:1302.6594

\bibitem[{{Zucker} {et~al.}(2004){Zucker}, {Kniazev}, {Bell},
  {Mart{\'{\i}}nez-Delgado}, {Grebel}, {Rix}, {Rockosi}, {Holtzman},
  {Walterbos}, {Annis}, {York}, {Ivezi{\'c}}, {Brinkmann}, {Brewington},
  {Harvanek}, {Hennessy}, {Kleinman}, {Krzesinski}, {Long}, {Newman}, {Nitta},
  \& {Snedden}}]{zucker04}
{Zucker}, D.~B., {Kniazev}, A.~Y., {Bell}, E.~F., {Mart{\'{\i}}nez-Delgado},
  D., {Grebel}, E.~K., {Rix}, H.-W., {Rockosi}, C.~M., {Holtzman}, J.~A.,
  {Walterbos}, R.~A.~M., {Annis}, J., {York}, D.~G., {Ivezi{\'c}}, {\v Z}.,
  {Brinkmann}, J., {Brewington}, H., {Harvanek}, M., {Hennessy}, G.,
  {Kleinman}, S.~J., {Krzesinski}, J., {Long}, D., {Newman}, P.~R., {Nitta},
  A., \& {Snedden}, S.~A. 2004, \apjl, 612, L121

\bibitem[{{Zucker} {et~al.}(2007){Zucker}, {Kniazev},
  {Mart{\'{\i}}nez-Delgado}, {Bell}, {Rix}, {Grebel}, {Holtzman}, {Walterbos},
  {Rockosi}, {York}, {Barentine}, {Brewington}, {Brinkmann}, {Harvanek},
  {Kleinman}, {Krzesinski}, {Long}, {Neilsen}, {Nitta}, \&
  {Snedden}}]{zucker07}
{Zucker}, D.~B., {Kniazev}, A.~Y., {Mart{\'{\i}}nez-Delgado}, D., {Bell},
  E.~F., {Rix}, H.-W., {Grebel}, E.~K., {Holtzman}, J.~A., {Walterbos},
  R.~A.~M., {Rockosi}, C.~M., {York}, D.~G., {Barentine}, J.~C., {Brewington},
  H., {Brinkmann}, J., {Harvanek}, M., {Kleinman}, S.~J., {Krzesinski}, J.,
  {Long}, D., {Neilsen}, Jr., E.~H., {Nitta}, A., \& {Snedden}, S.~A. 2007,
  \apjl, 659, L21

\end{thebibliography}

\bibliographystyle{apj}

\end{document}